\documentclass{JHEP3}
\usepackage{epsfig}

\newcommand{\be}{\begin{equation}}
\newcommand{\ee}{\end{equation}}
\newcommand{\bea}{\begin{eqnarray}}
\newcommand{\eea}{\end{eqnarray}}
 \newcommand{\IC}{\mathbb{C}}

\newcommand{\non}{\nonumber \\}

\def\IZ{\relax\ifmmode\hbox{Z\kern-.4em Z}\else{Z\kern-.4em Z}\fi}

 \def\hg{{\hat g}}

 \def\cn{{\cal N}}
 \def\co{{\cal O}}

\def\del{\partial}
 
\newcommand{\wt}[1]{\widetilde{#1}}

\def\del{{\partial}}

\def\al{\alpha}

\def\cn{{\cal N}} \def\cM{{\cal M}} \def\Mc{{\cal M}_c}
\newcommand{\sbsection}[1]{\vspace{.5cm} \noindent {\it #1}}

\preprint{{dated 3.3.03}}

\title{On Conformal Deformations II}

 \author{Barak Kol
\\
Racah Institute of Physics \\ The Hebrew University \\
Jerusalem 91904\\
Israel\\
\email{barak\_kol@phys.huji.ac.il} }

\abstract{The conformal index counts the number of exactly
marginal deformations. In 4d the index is given by the number of
chiral primary operators of dimension 3 moded out by the
complexified global group, where the quotient is defined as usual
by imposing a D-term. Here we show its consistency with the
Leigh-Strassler method for weakly coupled theories, and we test it
against known examples. In several examples this method discovers
extra exactly marginal deformations beyond those of
Leigh-Strassler. [This is an unpublished paper dated 3.3.03.]}


\begin{document}

\section*{Preface}

This is an unpublished paper dated 3.3.2003. \footnote{Files dated
by the computer operating system. This version is practically the
same as the original. Changes include: a new title page
 but no change in the abstract, added preface and new
acknowledgements, and a couple of minor non-scientific
proof-reading corrections.}
 Its purpose was to promote
\cite{KolIndex} both by refining the formalism and by analyzing
useful examples. \cite{KolIndex} introduced the role of the D-term
of the global group. Here it was refined in terms of the
holomorphic quotient into the central result, eq. (\ref{BKdim}).
 The paper ended up not being published because the logical step involved in the above-mentioned refinement was
well known and was judged not to merit publication.
\cite{KolIndex} presented the arguments which led me to this
result, and here it was further tested by many examples and found
to be precisely correct.
 Now I was extremely pleased that this central result was rigorously
derived by Green et. al. [arXiv:1005.3546].
%
Note that the superconformal index introduced in \cite{KolIndex}
(and further discussed here) can be considered to be realized by 
their derivation.

\section{Introduction}

Given a field theory, the first priority usually is to determine
the vacuum structure, and in particular the moduli space of vacua,
$\cM$ when one exists. Similarly, given a conformal field theory
we would like to know the conformal moduli space, $\Mc$, namely
the space generated by exactly marginal deformations. However,
whereas we know much about the moduli space of vacua following the
progress made in supersymmetric theories during the mid 90's, we
still know little about the conformal moduli space.

There are several properties of $\Mc$ which one would like to know
on the way to a complete solution: the dimension, the local
geometric structure (complex manifold, any non-trivial holonomy or
special geometry), and finally determination of the metric,
singularities and global issues. However, the dimension which is
simply the number of exactly marginal deformations is the only
topic which was studied so far.

Leigh and Strassler (1995) \cite{LS} discovered that exactly
marginal deformations are generic in 4d $\cn = 1$ supersymmetric
field theories and used exact $\cn = 1$ relations, including the
NSVZ formula \cite{NSVZ} to compute $\mbox{dim}(\Mc)$. In section
\ref{equiv} we will describe their result in full, but very
roughly it is \be
 \mbox{dim}(\Mc) = \#( \beta) - \# (\gamma ) \label{LSdim} \ee
Here $\# (\beta)$ is the number of supersymmetric marginal
operators, (or their associated couplings and $\beta$-functions),
and $\# (\gamma)$ is the number of anomalous dimensions
($\gamma$-functions) of the fundamental fields (including possibly
mixing). However, this ground breaking formulation had some
disadvantages: for a general CFT, not given by a Lagrangian the
concept of fundamental fields may not be well defined and so is
$\# (\gamma)$, moreover, the $\gamma$'s suffer from
non-gauge-invariance, and both $\beta$'s and $\gamma$'s are scheme
dependent.

In \cite{AKY} we studied the translation of this mechanism under
the AdS/CFT correspondence \cite{AdSCFT} 
 for the
case of 4d $\cn=4$. The translation is not straightforward since
both the $\beta$ and $\gamma$ functions do not have a well
understood translation. Following the supergravity analysis
 we claimed \cite{KolIndex} that in supergravity $\mbox{dim}(\Mc)$ is
given by the index of the supersymmetry variation operator, and
hence purely in field theory  \be \mbox{dim}
 (\Mc) = \mbox{Index}[\delta_{\mbox{superconf}}] \label{BKIndex} \ee
 where $\delta_{\mbox{superconf}}$ is the
superconformal variation operator considered to operate on the
space of operators. We defer the a direct discussion of the index
for later work and concentrate here on the more explicit
claim\footnote{\cite{KolIndex} included the main idea, but without
the full details. The relation between the $\gamma$ functions and
the complexified global transformation was discussed in
\cite{AKY}.} that locally at the origin of $\Mc$ \bea
 \Mc &\simeq& \mbox{supermarginals} / G_\IC \label{BKdim} \non
 \mbox{Definition: ``supermarginals '' } &\equiv&
\mbox{ chiral primary operators of dimension 3,}
\eea
 where  $G_\IC$
stands for the complexified global group, and the holomorphic
quotient is defined as usual by imposing the D-term (for the {\it
global} group) and then dividing by $G$.

Consequently, the generic dimension is \be
 \mbox{dim}(\Mc)=\mbox{dim}(R)-\left( \mbox{dim}(G)-\mbox{dim}(G_0) \right) \label{genericdim} \ee
 where $R$ denotes the vector space of supermarginals,
 $G_0 \subseteq G$ are global symmetries unbroken by any of the supermarginal couplings.
Interesting sub-generic cases do exist in which the dimension of
the holomorphic quotient is strictly smaller than the one above
(\ref{genericdim}). In this case our analysis shows that
(\ref{BKdim}) is correct to lowest order in the couplings, and
although in principle higher order contributions could change
that, this did not happen in several examples.

The merits of this ``index'' formulation are that it is valid for
any 4d $\cn=1$ CFT, all quantities are physical, and it requires
only knowledge of the chiral primary spectrum. The global group is
seen to play a central role on
$\Mc$ analogous to the gauge group on an $\cM$.

We should mention here two other motivations for the D-term. The
first is that when an AdS dual exists, the global symmetry group
becomes gauged and then the D-term is a necessary condition for a
susy vacuum (see \cite{Ceresole-DallAgata} for the
state-of-the-art on 5d supergravity). The second is that $\Mc$ is
a complex space,
 and the only way to perform a
quotient by the global group while keeping holomorphy is to impose
the D-term.

The relation (\ref{BKdim}) is at the center of the current paper.
Here it is precisely formulated\footnotemark[\value{footnote}],
 and it is confirmed and confronted against the LS formulation. In
section \ref{equiv} we show that when the LS formulation is valid
it coincides with (\ref{BKdim}), and that essentially they
computed the index for zero couplings. Then in section
\ref{examples} we re-analyze many of the examples of \cite{LS} and
some others. For each example we find the full set of exactly
marginal operators, which
 is often strictly larger than the ones found by \cite{LS}.
Table \ref{table1} summarizes the local description of $\Mc$ for
all the examples.


\section{An Equivalence with Leigh-Strassler}
\label{equiv}





\subsection{Set-up}

Let us start by setting up the notation for an arbitrary $\cn=1$
gauge theory. The local symmetry group is a product
$L=\prod_{i=1}^{n_L}\, L_i$. The fundamental matter fields
(generators of the chiral ring) are the chiral fields
$\phi_{s,t_s}$ where $s=1,\dots,n_R$ runs over the distinct
representations $R_s$ of $L$ and $t_s$ runs over
$t_s=1,\dots,T_s$, where $T_s$ is the multiplicity. Finally one
should specify a superpotential $W=W(\phi)$.

The classical global group (namely, no anomalies) for $W=0$ is
$\prod_{s=1}^{n_R}\, U(T_s)$.\footnote{A comment is due on the
$U(1)$ factors in the global group. One may choose to gauge an
anomaly-free $U(1)$, thereby removing it from the global group,
and at the same time constraining the possible operators that can
be added to the superpotential.} ``Holomorphic'' couplings in 4d
$\cn=1$, namely those which appear in that part of the Lagrangian
which is integrated over ``half of superspace'' ($\int\,
d^2\theta$),  are complex and may be divided into gauge couplings
$g_i,\, i=1,\dots,n_G$, \footnote{It is conventional to use the
complex coupling $\tau_I=\theta_I/(2\pi) + 4 \pi i/g_I^2$, where
$\theta_I$ is the associated theta angle, and $q_{g,I}= \exp(2 \pi
i\, \tau_I)$} and superpotential parameters $h_j$. In addition
there are the ``non-holomorphic'' couplings in the Kahler
potential, which we will not need, and henceforth ``couplings''
will denote the holomorphic ones unless stated otherwise.

The holomorphic marginal operators are generated from dimension 3
operators integrated over $d^2\theta$. The classical
supermarginals are the gauge coupling $g_i,\, i=1,\dots,n_L$ with
the associated operators $\mbox{Tr}(W_\al\, W^\al)$, and the
superpotential supermarginal, the coefficients of all dimension 3
gauge-invariant operators which may be added to $W$,  $h_j,\,
j=1,\dots, n_C$. Sometimes we shall refer to both $g_i$ and $h_j$
collectively as $\hg_k$.

\subsection{The Leigh-Strassler formulation}

Let us now describe the original formulation of Leigh-Strassler. A
set of couplings $\hg_k$ is exactly marginal if and only if all
their beta functions $\beta_k$ vanish (for all $\hg_k$).  $\cn=1$
imposes exact formulas for the $\beta$-functions in terms of the
$\gamma$-functions of the charged fields. Note that the
$\gamma$-functions of a specific representation $R_s$ are in the
adjoint of $U(T_s)$, since these fields can mix in the two point
function.

For a gauge coupling, $g$, the exact formula is the NSVZ formula
 \be \beta_{g_i} \sim f(g)\, [\beta_{0i} - \sum_{s=1}^{n_R}\, T(R_s,L_i)\, \gamma_s]
 \label{betag} \ee
where $\beta_0$ is the 1-loop beta function, $T_{s,i} \equiv
T(R_s,L_i)$ is the quadratic index of the representation
$\mbox{Tr}_{L_i,R_s} (T^A\, T^B) = T_{s,i}\, \delta^{A,B}$ and
$\gamma_s$ is the $U(1)$ component of the gamma functions. For the
beta function of a superpotential parameter $h$ ($\Delta W = h\,
\co$) the exact $\beta$ function is
 \be \beta_{h_j} \sim [\beta_0 + \sum_{s=1}^{n_R} \,  \gamma_s]\, h_j \label{betah} \ee
where $\beta_0 \propto -\Delta_W+\Delta_\co=-3+\Delta_\co$,
$\gamma_s$ is in the adjoint of $U(T_s)$, and hence the $U(1)$
charge of $h$ is $d(\phi_s,\co_j)={\del \log(\co_j) \over \del
\log(\phi_s)}$, namely the degree of $\phi_s$ in $\co_j$.

We concentrate on marginal couplings ($\beta_0=0$) and combine the
relations (\ref{betag},\ref{betah}) into \be
 0 = \beta_k \propto
\sum_{s=1}^{n_R}\, q_k^{s,A_s}\, \gamma_{s,A_s}\ee where $A_s$ is
an index in the adjoint of $U(T_s)$ and the matrix $q$ is given by
\be
\begin{array}{l||c||c|}
 & U(1)_1, \dots, U(1)_s, \dots, U(1)_{n_R} & SU(T_1), \dots,SU(T_s),\dots, SU(T_{n_R}) \\ \hline \hline
\begin{array}{c} g_i \\ i=1,\dots,n_L  \end{array} &  -T(R_s,L_i)  & -  \\ \hline \hline
\begin{array}{c} h_j \\ j=1,\dots,n_C  \end{array} &  d(\phi_s,\co_j) &  T^{A_s} \\
\hline  \end{array} \label{qmatrix} \ee

From this linear dependence it is deduced that in order to find
exactly marginals it is enough to set to zero all the
$\gamma$-functions \be
 \gamma_{s,A_s}=\gamma_{s,A_s}(g_i,h_j)=0, \label{LSeq} \ee
 or more precisely, it is enough to require the vanishing of
$\mbox{rank}(q)$ $\gamma$-functions which have non-zero $q$
coefficients. Generically there will be solutions when there are
fewer (independent) $\gamma$-functions than couplings, and the
generic dimension of the solution space is \be
 \dim{\Mc}=\#(\beta) - \mbox{rank}(q) \label{LSdimq} \ee

\subsection{Comparing with Leigh-Strassler}

We now compare our formulae (\ref{BKdim},\ref{genericdim}) with
(\ref{LSdimq}) at zero couplings. First, let us make the D-term
constraint, implicit in (\ref{BKdim}) explicit. The marginal
operators are of course gauge-invariant, but they may be charged
under the global group. The $h$'s have a standard charge given by
$d(\phi_s,\co_j)$, as in the q-matrix. We define the charges of
the $g$'s by using the q-matrix to be $T(R_s,L_i)$. Altogether the
D-term has the form \bea
 D^{A_s} &=& \sum_k\, \hg_k^\dagger\, T^{A_s}\, \hg_k = \sum_k\, \hg_k^\dagger\, q^{A_s}\, \hg_k = \non
 &=&  \sum_i\, T(R_s,L_i)\, \delta^{A_s,{\bf 1}}\,
 g_i^2- \sum_j\,  h_j^\dagger\, T^{A_s}\, h_j \label{Dterm} \eea
 where $\delta^{A_s,{\bf 1}}$ selects only the $U(1)$ generators
and the generator $T^{A_s}$ acts according to the representation
of $h$.

The couplings of LS are precisely the supermarginals for $W=g_i=0$
(when one of these conditions is violated some of the LS couplings
cease to be supermarginal as we will see soon). In order for the
two methods to agree it must be that the $\gamma$-function
equations coincide with the D-term equations to lowest order in
the coupling. First we should check that in both cases there is
the same number of equations. Indeed, since mixing in the
two-point-function is allowed only between fields in the same
representation, the $\gamma$-functions are valued in the global
group. So (\ref{LSdim}) can be rewritten as \bea
 \mbox{dim}(\Mc) &=& \#( \beta) - \# (\gamma ) \non
 &=& \#(\mbox{dim 3, chiral primaries for } W=0) - \mbox{dim(classical global group)}. \label{LSdim2} \eea
Moreover, picking only the independent $\gamma$-functions as in
(\ref{LSdimq}) is equivalent to correcting for the dimension of
$G_0$ the subgroup under which no coupling is charged, as in
(\ref{genericdim}), since when we turn on a coupling $\hg_k$ then
the violation of the global group generator $T^{A_s}$ is given by
the q-matrix $q_{j,A_s}$, and so dependent columns in $q$ generate
$G_0$.

\begin{figure}[t!]
\centering \noindent
\includegraphics[width=10cm]{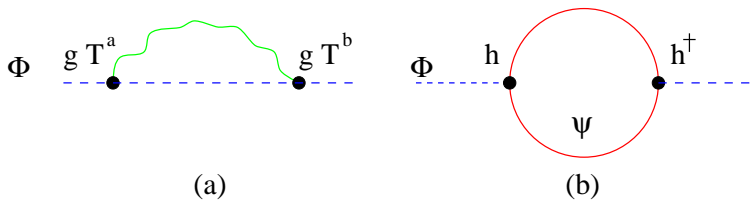}
\caption[]{Lowest order (1-loop) contribution to
the two point function of a field $\Phi$: (a) gauge boson
contribution (b) Fermion loop with a superpotential vertex. Note
that the diagrams have opposite signs due to the fermion loop.}
 \label{fig1}
\end{figure}

Next we would like to show that to lowest order $O(g^2,h^2)$ the
$\gamma$-functions coincide with the D-term. Looking at the 1-loop
contribution of the gauge bosons to the two-point-function, figure
\ref{fig1}(a), we see that it is indeed proportional to
$\mbox{Tr}_{L_i,R_s} (T^A\, T^B)\, g^2$.
 Similarly the $O(h^2)$ contribution is given by the
diagram in figure \ref{fig1}(b). The vertices come from
$\del_{s1,s2} W\, \psi_{s1}\, \psi_{s2}$ and its complex
conjugate, and the diagram is proportional to $- \sum_j\,
d(\phi_j,\co_i)\, h_j^\dagger\, T^{A_s}\, h_j$, where the minus
sign comes from the fermionic loop.


 Note that our ``consistency'' check at zero couplings is very close to a proof
that the two methods are identical, the only part missing being a
confirmation of all constants in the D-term constraint
(\ref{Dterm}).

Going beyond zero coupling, we need to study the effect of turning
on a coupling. The Konishi anomaly \cite{AmatiKonishiMeuriceRossi}
tells us that
 \be \{\bar{Q},\,
\bar{\psi}_{s,t_s}\, \phi_{s,u_s} \} = \sum_i T(R_s,L_i)\,
\mbox{tr}(\lambda_i\, \lambda_i) + {\del W \over \phi_{s,t_s}}\,
\phi_{s,u_s} \label{Konishi} \ee
 where $i=1,\dots,n_L$ runs over the
gauge groups, $s=1,\dots,n_R$ runs over the various matter
representations and $t_s,u_s=1,\dots,T_s$ where $T_s$ is the
representation multiplicity. Therefore whether the global group is
broken classically by $W$ or quantum mechanically by an instanton
(for $g>0$), for each global group generator which is broken on
the LHS of (\ref{Konishi}) the RHS ceases to be chiral primary
since it is expressed as $\bar{Q}(\mbox{something})$. Thus the
Konishi anomaly guarantees the index nature of (\ref{BKdim}) since
the difference in (\ref{genericdim}) stays constant as couplings
are turned on.

The argument above may continue to hold in some cases which are
not weakly coupled, but are still defined by a Lagrangian through
an RG flow. If one knows the fundamental fields (with non-trivial
anomalous dimensions this time) and the supermarginal operators,
then the exact formulae for the $\beta$-functions in terms of the
D-terms holds, only they are not homogeneous anymore since
$\beta_0=0$ does not necessarily hold. However, since we assume
that a fixed point does exist, then the dimension of the solution
space for the non-homogeneous problem reduces to that of the
homogeneous problem, which is the one we analyzed above.

Since the general claim (\ref{BKdim}) was shown to coincide with
Leigh-Strassler both at zero coupling and at small coupling, and
since it is phrased in terms of physically well-defined
quantities, we consider it to be plausible.

\subsection{Summary of method}

We summarize now our method (based on the principles in
\cite{KolIndex})

\begin{itemize}
 \item Identify the global group $G$, and the supermarginals. The
supermarginals consist of all gauge couplings and the dimension 3
gauge-invariant chiral primaries.
 \item Define the q-matrix of coupling charges as in
 (\ref{qmatrix}).
 \item Find the subgroup $G_0 \subseteq G$ under which no coupling
 is charged. This is equivalent to finding the number of dependent
 columns in the $U(1)$ part of the q-matrix.
 \item Determine the generic dimension to be (\ref{genericdim}).
 \item Continue to perform the holomorphic quotient (\ref{BKdim}) according to
 the D-term constraints (\ref{Dterm}), followed by a division by
 $G$. Often a $U(1)$ factor in $G$ can be cancelled against a
 supermarginal coupling, leading to simplification.
\end{itemize}

In the process of the imposing the D-terms it may happen that a
certain global $U(1)$ has only positive (gauge) or only negative
(superpotential) couplings. In that case all those couplings will
be forced to zero and the total dimension will be {\it
sub-generic}, at least to lowest order in the coupling. In
principle it could happen that higher order corrections lift this
degeneracy, but this did not happen is some examples: in an
$\cn=2$ orbifold of $\cn=4$ studied in \cite{AharonyRazamat} this
phenomena was seen first and it was shown to persist at least up
to third order; in a scalar theory (subsection \ref{scalar}) it is
known that there are no exactly marginals, to all orders.


\section{Examples}
\label{examples}












In this section we apply our method to a series of examples, most
of them from \cite{LS}. For each example we identify the global
group, $G$, and the ``supermarginal'' (chiral primaries of
dimension 3) operators, $R$ (at the origin of $\Mc$). The set of
exactly marginal operators is then given by the holomorphic
quotient $R/ G_\IC$. The exactly marginal operators of \cite{LS}
were found by applying intuition and special discrete symmetries
to solve their equations, but were not claimed to exhaust the
whole set of exactly marginals, and indeed we often find a
strictly larger set, the first case being subsection \ref{Nc3}.
Table \ref{table1} summarizes the local description of $\Mc$ for
all the examples.



\TABLE{$
   \begin{array}{|l|c|c|} \hline
 \mbox{Theory} & \mbox{Global group, } G & \mbox{supermarginals, } R \\
 \hline \hline
 \cn=4                              & SU(3)   & {\bf 10}+{\bf 1}          \\ \hline
 \cn=2 \mbox { w. } A,\, S          & U(1)^2      & {\bf 1}_{2,0} + {\bf 1}_{-2,0}+{\bf 1}_{0,0} \\
 \cn=2 \mbox { w. } N_f=2\, N_c & SU(N_f)_{\mbox{diag}} \times U(1)_B & {\bf 1_0} \\ \hline
 \mbox{SQCD w. } N_c=3, ~N_f=9    & SU(9)_L \times SU(9)_R
   \times U(1)_B    & {\bf (84,1)_3} + {\bf (1,{\bf 84})_{-3}} \\ \hline
 \mbox{SQCD w. } N_c/N_f=1/2        & SU(N_f)_L \times SU(N_f)_R
   \times U(1)_B    & S^2({\bf (N_f,N_f)})_0 \\
   \mbox {same w. }N_c = 4 & &
  \begin{array}{c} \mbox{same }+{\bf (70,1)_4} + {\bf (1,70)_{-4}} \\ \mbox{(baryons)} \\ \end{array} \\ \hline
 \mbox{scalar theory} & \multicolumn{2}{c|}{\mbox{no exactly marginals due to sub-generic
 D-term}}   \\ \hline
 \begin{array}{l} SU(N) \times SU(N) \\ \mbox{ w. } 3[{\bf (N,\bar{N})}+{\bf (\bar{N},N)}] \\ \end{array} &  SU(3) \times SU(3) \times
 U(1)_V & \mbox{no exactly marginals}  \\
 \mbox {same w. }N = 3 &  & {\bf 1}+{\bf (10,1)_3} + {\bf (1,10)_{-3}} \\
 \hline \hline
 SU(2) \times SU(2) \mbox{ w. } {\bf (3,3)} & - & {\bf 1}+{\bf 1} \\ \hline
 E(6) \mbox{ w. } 12 \cdot {\bf 27}       & SU(12) & S^3({\bf 12})={\bf 182} \\ \hline
 SU(4) \mbox{ w. } 8\, ({\bf 4} + {\bf \bar{4}}) + 4 \cdot {\bf 6} &
 SU(8)_Q \times SU(8)_{\wt{Q}} \times SU(4)_A \times U(1)^2 & {\bf (28,1,4)_{2,0}} + {\bf (1,28,4)_{-2,0}}  \\ \hline
\end{array} $\caption{Summary of examples and results. For each theory we state the
global group$^*$, and the supermarginal operators -- the dimension
3 chiral primaries (in a complex representation). $g>0$ is assumed
when relevant. Locally around the origin $\Mc$ is given by $\Mc
\simeq R/ G_\IC$ -- note that $G$ ``plays the role'' of a local
group on $\Mc$ and the appearance of a D-term in the quotient.
*The specified global group does not include a $U(1)_R$ factor. In
cases with extended supersymmetry and higher R-symmetry, only the
commutant with $U(1)_R$ is given.} \label{table1} }

\subsection{$\cn=4$}

We worked out this case in detail in \cite{AKY} and we include it
here for completeness. In an $\cn=1$ language the matter content
is three adjoints, and the superpotetial is $W_0=g\, \mbox{tr}\,
([\phi_1,\, \phi_2]\, \phi_3)$. The global group is $SO(6)_R$ out
of which $SU(3) \times U(1)_R$ is manifest in $\cn=1$ language.

For any simple gauge group the theory has one exactly conformal
parameter which preserves $\cn=4$, namely the complex gauge
coupling
 \footnote{In particular the one loop beta function vanishes $b_0 \propto 3-3=0$.}.
  For some gauge groups cubic
invariants exist, in particular for $SU(N)$ one has the symmetric
invariant $d^{ABC}=\mbox{Tr}\, (\{ T^A,\, T^B \} T^C)$ where $T^A$
are group generators. Therefore $S^3({\bf 3})={\bf 10}$ of $SU(3)$
 are supermarginals. In addition, for zero coupling $W_0$ is
supermarginal as well. The $q$-matrix is \be
\begin{array}{l||c||c|}
    & U(1)_\phi & SU(3)_\phi \\
    \hline \hline
 g  & -3\, N         &  - \\ \hline \hline
 W_0 & 3 & {\bf 1} \\ \hline
  \phi^3    & 3 & {\bf 10} \\ \hline
\end{array} \ee
 A combination of $g$ and $W_0$ produces the fixed line of $\cn=4$
 and the q-matrix reduces to \be
\begin{array}{l||c|}
    &  SU(3)_\phi \\
    \hline \hline
 g\, \& \, W_0  & {\bf 1} \\ \hline \hline
  \phi^3    & {\bf 10} \\ \hline
\end{array} \ee

Hence \be
 \Mc \simeq {\bf 10}/SL(3,\IC) + {\bf 1}_\IC . \ee

In the zero coupling limit the global group enhances $SU(3)_\phi
\to U(3)_\phi$ and this effect in cancelled by a new
supermarginal, namely $W_0$, which is allowed since $W=0$.

In \cite{LS} IV.D two exactly marginal deformations are
demonstrated $\phi_1\, \phi_2\, \phi_3$ and $\phi_1^{~3} +
\phi_2^{~3} + \phi_3^{~3}$ based on some special discrete subgroup
of $SU(3)$ under which these operators are invariant. Of course
these operators could be rotated by $SU(3)$ so we should think of
them as being representatives. Since \cite{LS} finds here
$2_\IC=10-8$ operators they exhaust all exactly marginals, except
for the subtlety that the $SU(3)$ action actuallly induces some
discrete identification by some finite subgroup of $SU(3)$ (see
\cite{AKY}, appendix).

\subsection{$\cn =2$}

Take an $\cn=2$ theory with matter in the symmetric and
anti-symmetric. In $\cn=1$ language an $\cn=2$ matter
hypermultiplet in representation $R$ doubles into chiral
multiplets in $R$ and in its complex conjugate $\widetilde{R}$,
and in this case we have matter in $ S,\, \widetilde{S},\, A,\,
\widetilde{A}$. In addition the $\cn=2$ vector multiplet
contributes an adjoint $\phi$. The superpotential is of the form
$W_0=g\, (\widetilde{A}\, \phi\, A + \widetilde{S}\, \phi\, S)$.

At zero couplings the global symmetry is $U(1)^5 = U(1)_S \times
U(1)_{\widetilde{S}} \times U(1)_A \times U(1)_{\widetilde{A}}
\times U(1)_\phi$
 \footnote{Actually $U(1)_\phi$ should be replaced by the $\cn=2$ $SU(2)_R$
 which is not manifest in $\cn=1$.}.
 The supermarginals are the gauge coupling
($b_0=3N-N_\phi-(N+2)_S-(N-2)_A=0$), and the five dimension 3
operators $\phi^3,\, \wt{S}\, \phi\, S,\, \wt{A}\, \phi\, S,\,
\wt{S}\, \phi\, A,\, \wt{A}\, \phi\, A$, and the q-matrix is
\be \begin{array}{l||c|c|c|c|c|}
   & U(1)_S & U(1)_{\wt{S}} & U(1)_A & U(1)_{\wt{A}} & U(1)_\phi \\ \hline \hline
 g & -(N+2)/2 & -(N+2)/2 &  -(N-2)/2 & -(N-2)/2      & -N \\ \hline \hline
\wt{S}\, \phi \, S & 1 & 1 & 0 & 0 & 1 \\ \hline
\wt{A}\, \phi \, A & 0 & 0 & 1 & 1 & 1 \\ \hline
\wt{S}\, \phi \, A & 0 & 1 & 1 & 0 & 1 \\ \hline
\wt{A}\, \phi \, S & 1 & 0 & 0 & 1 & 1 \\ \hline
\phi^3             & 0 & 0 & 0 & 0 & 3 \\ \hline
\end{array} \ee 
    One notices that there is one combination which is preserved by all supermarginals, namely, $G_0 = U(1)_{(S-\wt{S})-(A-\wt{A})}$. Therefore the (generic) dimension is $6-4=2_\IC$.

Let us study this space in more detail. We can eliminate one coupling, and one global $U(1)$ by noticing that the D-term equation for $U(1)_{\phi-(S+\wt{S}+A+\wt{A})}$ is simply $h_{\phi^3}=0$. Next, there is one exactly marginal (the $\cn=2$ coupling) which is a mixture of $g,\,, \wt{S}\, \phi \, S$ and $\wt{A}\, \phi \, A$. This can be seen by looking at the appropriate rows in the q-matrix, and noticing that for these rows the reduced q-matrix degenerates as $U(1)_S=U(1)_{\wt{S}},\, U(1)_A=U(1)_{\wt{A}}$ and hence there are 2 equations for 3 couplings. Finally, on the $\cn=2$ fixed line the global symmetry is $U(1)_{S-\wt{S}} \times U(1)_{A-\wt{A}}$ and there are two additional supermarginals $\wt{S}\, \phi \, A,\, \wt{A}\, \phi \, S$. Since there is the abovementioned combination of the global $U(1)$'s which preserves both, we get a second exact marginal.

All in all we find the same exact marginals as in \cite{LS} section IV.E.

Another conformal $\cn=2$ example is given by a $\cn=2$ SQCD with
$N_f= 2\, N_c$. This example was treated by Seiberg-Witten
\cite{SeibergWitten2} (and not in \cite{LS}), and is known to have
$\mbox{dim}(\Mc)=1_\IC$ namely the complex gauge coupling.

Let us confirm this result using our formulation. In terms of
$\cn=1$ at zero coupling we have the q-matrix
\be \begin{array}{l||c|c|c||c|c|}
  & U(1)_L & U(1)_R & U(1)_\phi & SU(N_f)_L & SU(N_f)_R \\ \hline \hline
g & -N_c   & -N_c   & -N_c      & -         & -  \\ \hline \hline
\wt{Q}\, \phi\, Q & 1 & 1 & 1   & {\bf N_f}  & {\bf N_f} \\ \hline
\phi^3 & 0 & 0 & 3 & {\bf 1} & {\bf 1}  \\ \hline
\end{array} \ee

As in the previous example $U(1)_\phi$ and $\phi^3$ can be both eliminated after considering the D-term equation for $U(1)_{\phi-(R+L)/2}$. Next we notice that $U(1)_B \equiv U(1)_{R-L}$ is preserved by all supermarginals. Hence the q-matrix reduces to
 \be \begin{array}{l||c||c|c|}
    & U(1)_{L-R} & SU(N_f)_L & SU(N_f)_R \\ \hline \hline
 g  & -2\, N_c  & -  & -  \\ \hline \hline
 \wt{Q}\, \phi\, Q & 2    & {\bf N_f}  & {\bf N_f} \\ \hline
 \end{array} \ee
 Performing the quotient $({\bf N_f},{\bf N_f})/(SL(N_f,\IC) \times SL(N_f,\IC)$ a singlet remains, and then the D-term for the $U(1)$ determines $g$.

 Alterntively, the analysis can be repeated on the $\cn=2$ fixed line. Now the global group is broken down to $SU(N_f)_{\mbox{diag}} \times U(1)_B$ (the axial $U(1)$ is broken by instantons). The supermarginals  are in $(\mbox{\textbf{adjoint}} + {\bf 1})_0$, and so after the division by $G$ only the singlet remains.

\subsection{SQCD with $N_c=3 , ~N_f=9$}
\label{Nc3}

This is our first example without extended supersymmetry, and the first where we display more exact marginals than in \cite{LS}. The
number of colors is especially tuned to allow for cubic
invariants.

The matter content of 9 flavors guarantees $b_0=3\, N_c-N_f = 0$
and generates a classical global group $U(1)_Q \times U(1)_{\wt{Q}}
\times SU(9)_Q \times SU(9)_{\wt{Q}}$. The supermarginals are $g$ and the gauge invariant cubics, namely
$Q^3$ and $\wt{Q}^3$. The q-matrix is \be
 \begin{array}{l||c|c||c|c|}
    & U(1)_Q & U(1)_{\wt{Q}}& SU(9)_Q    & SU(9)_{\wt{Q}} \\
    \hline \hline
 g  & -9/2    & -9/2        &  -  & - \\ \hline \hline
 Q^3 & 3 & 0    & S^3({\bf 9})={\bf 84} & {\bf 1} \\ \hline
 \wt{Q}^3 & 0 & 3   & {\bf 1} &   {\bf 84} \\ \hline
\end{array} \ee
 Instantons break the $U(1)^2$
part into $U(1)_B$ whose charges are defined by $q_B=q_Q -
q_{\wt{Q}}$. After eliminating $g$ against the broken axial $U(1)$ the q-matrix reduces to
  \be \begin{array}{l||c||c|c|}
    & U(1)_B & SU(9)_Q    & SU(9)_{\wt{Q}} \\
    \hline \hline
 Q^3 & 3     & S^3({\bf 9})={\bf 84} & {\bf 1} \\ \hline
 \wt{Q}^3 & -3    & {\bf 1} &  {\bf 84} \\ \hline
\end{array} \ee

\subsection{SQCD with $N_c/N_f=1/2$}

In this example we find a large $\Mc$ for one of the simplest (and
relatively physical) theories.

SQCD within the ``conformal window'' $1/3<N_c/N_f <2/3$ flows in
the IR to a non-trivial CFT. The dimension of the meson fields
$M=Q\, \wt{Q}$ is given by $D(M)= 3\,(1- N_c/ N_f)$. If one wants
to construct supermarginal operators, it is necessary to choose
$N_c/N_f=1/2$ and hence \be D(M)= 3/2 \ee

The global symmetry is known to be $U(1)_B \times SU(N_f)_Q \times
SU(N_f)_{\wt{Q}}$ (the axial $U(1)$ is broken by the instantons).

The supermarginals are of the form $M^2=Q^2\, \wt{Q}^2$ and are in
the representation $S^2({\bf N_f},{\bf N_f})_0=({\bf
S^2(N_f)},{\bf S^2(N_f)})_0+({\bf A^2(N_f)},{\bf A^2(N_f)})_0$.

The q-matrix is simply \be \begin{array}{l||c||c|c|}
    & U(1)_B    & SU(N_f)_Q & SU(N_f)_{\wt{Q}} \\ \hline \hline
M^2 & 0         & {\bf S^2(N_f)} & {\bf S^2(N_f)} \\
    & 0         & {\bf A^2(N_f)} & {\bf A^2(N_f)} \\ \hline
\end{array} \ee
and so \be
 \Mc \simeq S^2({\bf N_f},{\bf N_f}) /SL(N_f,\IC) \times SL(N_f,\IC) \ee
 which has dimension $N_f^{~2}\, (N_f^{~2}-3)/2+2$.

For the special case $N_c=4, \, N_f=8$ there are additional
baryonic supermarginals, $Q^4$ and  $\wt{Q}^4$. They lie in
representation $A^4({\bf 8})_4 + \mbox{c.c.} ={\bf 70}_4 + {\bf
70}_{-4}$. Now the D-term constraint for $U(1)_B$ is not trivial
anymore and should be added.

\subsection{Scalar theory}
\label{scalar}

It is well known that theories with no gauge fields do not have
exactly marginals \cite{ColemanGross}, and actually any marginal
perturbation leads to a Landau pole in the UV. From our point of
view this is a non-generic case where the dimension formula
(\ref{genericdim}) and the quotient formula (\ref{BKdim})
disagree, as a single D-term equation can impose the vanishing of
several couplings. In this example we see that the quotient
formula is correct.

The simplest example is a theory with 2 scalars $A_i,\, i=1,2$.
The q-matrix is given by \be \begin{array}{l||c||c|}
    & U(1)_A & SU(2)_A \\ \hline \hline
A^3 &   3 & {\bf 4} \\ \hline
 \end{array} \ee

Although there are 4 supermarginals (parametrized by $h_1\, A^3 +
h_2\, A^2\, B + h_3\, A\, B^2 + h_4\, B^3$) and only 2 global
$U(1)$'s, there are no exact marginals as can be seen from the
1-loop $\gamma$-functions \bea
 \gamma_A = 3\, |h_1|^2 + 2\, |h_2|^2 +  |h_3|^2 = 0 \\
 \gamma_B = |h_2|^2 + 2\, |h_3|^2 + 3\, |h_4|^2 = 0 \\
 \eea
 These two equations force all $h$'s to vanish.
One could be concerned whether higher order contributions to the
$\gamma$'s could change the picture, such as adding $-|h_4|^4$ to
$\gamma_A$, but then the general theorem \cite{ColemanGross}
forbids such a correction.

\subsection{$SU(N_c) \times SU(N_c)$ with 3 bifundamentals}

This example shows that you cannot get exactly marginals starting with only gauge couplings.

The theory has gauge group $SU(N_c) \times SU(N_c)$ and matter in
$3\, [Q + \wt{Q}] \equiv 3\, [({\bf N_c},{\bf \bar{N_c}})+({\bf
\bar{N_c}},{\bf N_c})]$. Each factor of the gauge group has
effectively $N_f=3\, N_c$ and so $b_0=0$. The classical global
group is $U(3)_Q \times U(3)_{\wt{Q}}$. For generic $N_c$ the only
supermarginals are the gauge couplings and the q-matrix is \be
\begin{array}{l||c|c||c|c|}
    & U(1)_Q  & U(1)_{\wt{Q}} & SU(3)_Q  & SU(3)_{\wt{Q}} \\ \hline \hline
 g_1  & -3\, N_c/2 & -3\, N_c/2 & - & - \\ \hline
 g_2  & -3\, N_c/2 & -3\, N_c/2 & - & - \\ \hline
 \end{array} \ee

We see that the q-matrix is degenerate, the $U(1)_B$ combination
is unbroken, and we change basis to $U(1)_B,\, U(1)_{Q+\wt{Q}}$.
The q-matrix becomes \be \begin{array}{l||c|c||c|c|}
    & U(1)_{Q+\wt{Q}} & U(1)_B & SU(3)_Q  & SU(3)_{\wt{Q}} \\ \hline \hline
 g_1  & -3\, N_c & 0 & - & - \\ \hline
 g_2  & -3\, N_c & 0 & - & - \\ \hline
 \end{array} \ee

At first it looks like due to the degeneracy of the q-matrix we
are left with one constraint on two couplings. However, this
D-term constraint has only negative charges $0=-3\, N_c\, g_1^{~2}
-3\, N_c\, g_2^{~2}$ and hence there is no solution (this result
is robust against corrections since the quadratic form $g_1^{~2}
+g_2^{~2}$ is non-degenerate and thus the local behavior will not
be changed by higher order corrections). This is not surprising as
we expect the balance of $g$'s and $h$'s to be necessary.

For $N_c=3$ $h$'s can be formed, and $\mbox{dim}(\Mc)$ returns to
be generic. The additional supermarginals are $Q^3,\, \wt{Q}^3$ in
the $({\bf 10},{\bf 1})_3,\, ({\bf 1},{\bf 10})_{-3}$. The
q-matrix is \be \begin{array}{l||c|c||c|c|}
    & U(1)_{Q+\wt{Q}} & U(1)_B  & SU(3)_Q  & SU(3)_{\wt{Q}} \\ \hline \hline
 g_1  & -3\, N_c & 0 & - & - \\ \hline
 g_2  & -3\, N_c & 0 & - & - \\ \hline \hline
 Q^3 & 3 & 3 & {\bf 10} & {\bf 1} \\ \hline
 \wt{Q}^3 & 3 & -3 & {\bf 1} & {\bf 10} \\ \hline
 \end{array} \ee
 The non-Abelian quotient gives two ${\bf 10}/SL(3,\IC)$ factors, the $U(1)_B$ D-term
equates the absolute values of these two factors, and finally the
$U(1)_{Q+\wt{Q}}$ D-term gives
$N_c(g_1^{~2}+g_2^{~2})=|h_{Q^3}|^2+|h_{\wt{Q}^3}|^2$ so that only
the sum $g_1^{~2}+g_2^{~2}$ is fixed and thus an additional
exactly marginal is reclaimed. Thus altogether we get \be [{\bf
10}_3/SL(3,\IC) +{\bf 10}_{-3}/SL(3,\IC)]/\IC^*+{\bf 1} \ee where
the last ${\bf 1}$ arises from the gauge couplings.

\subsection{Others}

Here we discuss and compare several other cases that appeared in \cite{LS}.

\sbsection{$SU(2) \times SU(2)$ with ({\bf 3},{\bf 3})}

The matter is denoted by $ Q \equiv ({\bf 3},{\bf 3})$. For each
gauge group factor we have effectively 3 adjoints and so $b_0=0$.
The classical global group is $U(1)_Q$, and the supermarginals are
$g$ and $Q^3$. The resulting q-matrix is \be \begin{array}{l||c|}
    & U(1)_Q \\ \hline \hline
  g_1 & -3 \cdot 2 \\ \hline
  g_2 & -3 \cdot 2 \\ \hline \hline
  Q^3 & 3 \\ \hline \end{array} \ee
The instantons break the $U(1)_Q$ and we are left with two exactly
marginals (one of them gauge like in the previous example?) Here
our results agree with \cite{LS} section IV.B.

\sbsection{$E_6$ with 12\, {\bf 27}}

This example was brought in \cite{LS} as an example with chiral
matter. From our perspective the gauge group and its chiral
representations do not play a role, and the only difference is
that the global group will not include factors for both $Q$
and $\wt{Q}$.

The matter content was chosen so that $b_0=3\, T({\bf 78})-12\,
T({\bf 27})=3 \cdot 24 -12 \cdot 6=0$. The classical global group
is $U(12)_Q$ broken by instantons to $SU(12)_Q$. In order to form
cubics in $Q$ it is important that $E_6$ has an invariant cubic
symmetric tensor (for the ${\bf 27}$), and so $Q^3$ sits in
$S^3({\bf 12})_3 = {\bf 182}_3$. Altogether we find \be \Mc \simeq
{\bf 182}/SL(12,\IC) \ee This $39_\IC$ dimensional space is much
larger than the special cases found in \cite{LS} section IV.F.

\sbsection{$SU(4)$ with 8 flavors and 4 antisymmetrics}

This theory is brought up in \cite{LS} as an example with a global
$U(1)$ which is unbroken by instantons (``undetermined R
charge''). For us this is not a new feature, but it is an interesting
application for our method.

The matter content satisfies $b_0=3\, N_c - N_f - 4\, T(A) =
12-8-4=0$, where $T(A)=(N-2)/2=1$. The classical global group is
$U(8)_Q \times U(8)_{\wt{Q}} \times U(4)_A$ and the supermarginals
are the gauge coupling and $Q^2\, A,\, \wt{Q}^2\, A$. The q-matrix
is given by \be \begin{array}{l|c|c|c||c|c|c|}
    & U(1)_Q & U(1)_{\wt{Q}} & U(1)_A & SU(8)_Q & SU(8)_{\wt{Q}} &
    SU(4) \\ \hline \hline
g   & -4    & -4    & -4 & - & - & - \\ \hline \hline
 Q^2\, A & 2 & 0 & 1 & A^2({\bf 8})={\bf 28} & {\bf 1} & {\bf 4} \\
 \hline
 \wt{Q}^2\, A & 0 & 2 & 1 & {\bf 1} & {\bf 28} & {\bf 4} \\
 \hline \end{array} \ee

Out of the $U(1)^3$ two are unbroken by instantons, and can be
chosen to be $U(1)_{2A-Q-\wt{Q}}$ (which is preserved by all
supermarginals) and $U(1)_B=U(1)_{Q-\wt{Q}}$). The reduced
q-matrix becomes \be \begin{array}{l|c|c|c||c|c|c|}
    & U(1)_B & U(1)_{2A-Q-\wt{Q}} &  SU(8)_Q & SU(8)_{\wt{Q}} &
    SU(4) \\ \hline \hline
 Q^2\, A & 2 &  0 & A^2({\bf 8})={\bf 28} & {\bf 1} & {\bf 4} \\
 \hline
 \wt{Q}^2\, A & -2 & 0 & {\bf 1} & {\bf 28} & {\bf 4} \\
 \hline \end{array} \ee from which we deduce
 \be \Mc \simeq \left[ ({\bf 28},{\bf 4})_2/SL(8,\IC) + ({\bf 28},{\bf
 4})_{-2}/SL(8,\IC) \right] / \left( SL(4,\IC) \times \IC^*
 \right). \ee

\vspace{0.5cm} \noindent {\bf Acknowledgements}

It a pleasure to thank Ofer Aharony for collaboration in early stages of the work and for many enjoyable discussions.
For additional discussions I would like to thank A. Armoni, V. Asnin, S. Cherkis, A. Giveon, A. Hanany, A. Konnechny and Y. Oz.


In 2003 the work was partially supported by The Israel Science
Foundation grant no 228/02. At present (2010) I am supported by The Israel Science Foundation grant
no 607/05, by the German Israel Cooperation Project grant DIP H.52, and the Einstein Center at the Hebrew University.


\end{document}